%% file: main.tex
\documentclass[conference]{IEEEtran}
\usepackage{fancyhdr}
\IEEEoverridecommandlockouts

\usepackage[english]{babel}
\usepackage{amsmath,amssymb,amsfonts}
\usepackage{newtxtext,newtxmath}
\usepackage{booktabs}
\usepackage{cite}
\usepackage{enumitem}
\usepackage{graphicx}
\usepackage{float}
\usepackage{textcomp}
\usepackage{xurl}
\usepackage[hidelinks]{hyperref}
\usepackage{makecell}
\title{Provisioning Networks for AI Supercomputers:\\
A Trace-Driven Study of Performance Sensitivity at Unprecedented Scale}

\makeatletter
\newcommand{\linebreakand}{%
  \end{@IEEEauthorhalign}
  \hfill\mbox{}\par
  \mbox{}\hfill\begin{@IEEEauthorhalign}}
\makeatother

\author{
\IEEEauthorblockN{Tommaso Bonato\textsuperscript{1}}
\IEEEauthorblockA{ETH Zurich; Microsoft Corporation\\
Zurich, Switzerland; Redmond, USA\\
tommaso.bonato@inf.ethz.ch}
\and
\IEEEauthorblockN{Amirreza Rastegari}
\IEEEauthorblockA{Microsoft Corporation\\
Redmond, USA\\
arastegari@microsoft.com}
\and
\IEEEauthorblockN{Daniele De Sensi}
\IEEEauthorblockA{Sapienza University of Rome\\
Rome, Italy\\
desensi@di.uniroma1.it}
\and
\IEEEauthorblockN{Timo Schneider}
\IEEEauthorblockA{ETH Zurich\\
Zurich, Switzerland\\
timo.schneider@inf.ethz.ch}
\linebreakand
\IEEEauthorblockN{Siyuan Shen}
\IEEEauthorblockA{ETH Zurich\\
Zurich, Switzerland\\
siyuan.shen@inf.ethz.ch}
\and
\IEEEauthorblockN{Ales Kubicek}
\IEEEauthorblockA{ETH Zurich\\
Zurich, Switzerland\\
akubicek@student.ethz.ch}
\and
\IEEEauthorblockN{Prabhat Ram\textsuperscript{1}}
\IEEEauthorblockA{Microsoft Corporation\\
Redmond, USA\\
prabhat.ram@microsoft.com}
\and
\IEEEauthorblockN{Torsten Hoefler}
\IEEEauthorblockA{ETH Zurich\\
Zurich, Switzerland\\
torsten.hoefler@inf.ethz.ch}
}

\begin{document}

\maketitle

\thispagestyle{fancy}
\lhead{}
\rhead{}
\chead{}
\lfoot{\footnotesize{%
SC26, November 15-20, 2026, Chicago, Illinois, USA
\newline 979-8-3195-4789-7/26/\$31.00 \copyright\ 2026 IEEE}}
\rfoot{}
\cfoot{}
\renewcommand{\headrulewidth}{0pt}
\renewcommand{\footrulewidth}{0pt}

\begingroup
\renewcommand{\thefootnote}{\arabic{footnote}}
\footnotetext[1]{Tommaso Bonato and Prabhat Ram have since left Microsoft Corporation.}
\endgroup
\setcounter{footnote}{1}

\input{Sources/abstract}
\input{Sources/intro}
\input{Sources/motivation}
\input{Sources/impl}
\input{Sources/evaluation}
\input{Sources/related}
\input{Sources/conclusions}

\bibliographystyle{IEEEtran}
\bibliography{references}

\end{document}

%% file: Sources/abstract.tex
\begin{abstract}\label{sec:abstract}
AI workloads on supercomputers make interconnect performance a primary design constraint. Hardware specifications and microbenchmarks alone do not show how real workloads respond to latency, bandwidth, and jitter. We present a trace-driven study of network sensitivity for training and inference workloads at scales up to 4,096 GPUs. Our methodology combines signature-based trace decomposition with linear programming to make broad network-parameter sweeps feasible. It preserves the recorded NCCL schedules and program dependencies, and estimates a lower bound on runtime under the LogGP model. We validate the compositional analysis against monolithic linear programs, LogGOPSim, and hardware measurements at feasible scales. The resulting latency-bandwidth envelopes identify low-sensitivity regions and performance cliffs that vary substantially across workloads. They also show when bandwidth upgrades provide meaningful returns and when additional network performance has little effect. These results translate execution traces into practical guidance for evaluating network configurations while making the model assumptions explicit.
\end{abstract}


%% file: Sources/intro.tex
\section{Introduction}

The scale of AI training infrastructure has grown by orders of magnitude in less than five years~\cite{meta_ai_center_wagner, cottier_ml_costs, cscs_alps}. Clusters built to train frontier models now contain tens of thousands of accelerators interconnected by networks whose aggregate bandwidth exceeds petabits per second~\cite{hpn_alibaba, hammingmesh_hoefler}. The capital expenditure for a single such facility can easily exceed one billion dollars, and roughly 20\% of the cost of training frontier models is attributable to the network fabric~\cite{cottier_ml_costs}. At this investment scale, the question of how much network performance AI workloads actually need is not merely academic. Overprovisioning wastes capital. Underprovisioning degrades training throughput, increasing time to convergence and inflating operational costs.

\begin{figure}[!t]
\centering
\includegraphics[width=1\linewidth]{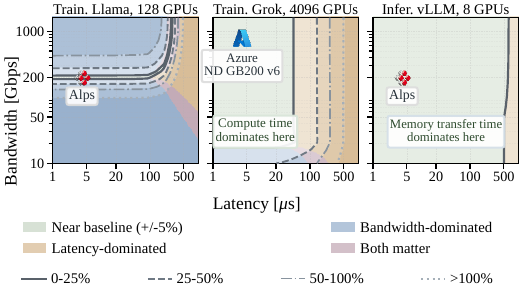}
\caption{Latency/bandwidth sensitivity maps constructed from separate model-derived sweeps for each AI workload. Colors indicate the dominant runtime sensitivity regime; contours show increasing runtime change relative to each system's logo-marked baseline operating point.}
\label{fig:poster}
\end{figure}

Despite the stakes, the trade-off between network and model performance is not well studied. The dominant approach to AI cluster design treats network performance as a constraint to be maximized within a power and cost envelope: more bandwidth is better, less latency is better. This approach is reasonable when the relationship between network parameters and application performance is unknown, but it can lead to overprovisioning for some models and parallelism strategies. As shown in Figure~\ref{fig:poster}, obtained with the toolchain presented in Section~\ref{sec:design}, workloads such as xAI's Grok-1~\cite{grok1_xai} and vLLM~\cite{vllm_kwon} with Llama~8B~\cite{llama3_grattafiori} exhibit large regions of the latency-bandwidth parameter space that produce nearly identical training and inference runtimes under our model. Llama~70B, by contrast, is predominantly bandwidth-bound and becomes latency-sensitive only at very high latency values.

Historically, studying this trade-off has been difficult for multiple reasons: First, AI training workloads exhibit communication patterns that are far more complex than those of traditional HPC applications. A single training iteration of a large language model (LLM) interleaves computation with collective communication across multiple CUDA streams, using a mix of tensor, pipeline, data, and expert parallelism~\cite{megatron_shoeybi, gpipe_huang, zero_rajbhandari, deepseek_v3}. The resulting execution graph is a dense directed acyclic graph (DAG) with millions of vertices whose critical path depends on the network parameters in non-obvious ways. Second, the scale of production training (over tens of thousands of GPUs) makes direct experimentation prohibitively expensive. Running a single training job for long enough to obtain stable throughput measurements at one network configuration consumes thousands of GPU-hours; sweeping over a meaningful parameter grid is infeasible. Third, packet-level network simulators such as NS-3~\cite{ns3_riley} or htsim~\cite{htsim_raiciu}, while accurate, are too slow to serve as the basis for systematic parameter studies at scale~\cite{towards_million_besta, atlahs_bonato, llamp_shen}.

This question also cannot be answered from microbenchmarks alone. Real training executions combine communication and computation in dependency-constrained phases whose performance depends on temporal ordering, overlap, burstiness, and synchronization structure~\cite{atlahs_bonato}. Two workloads with similar aggregate communication volume may expose different critical paths once these effects are accounted for. Even within one workload, the relevant network bottleneck can shift across phases and scale points. Synthetic kernels remain valuable controls, but they do not capture the complete behavior of end-to-end AI workloads.

On the other side, small-scale hardware experiments are expensive and do not always tell the full large-scale story. As jobs scale, the balance between exposed communication and useful computation changes, collective algorithms switch regimes (from being communication-bound to computation-bound and vice versa), and synchronization amplifies small service differences into iteration-level slowdowns~\cite{noise_in_the_clouds_de_sensi, dean_tail_at_scale}. The resulting sensitivity landscape is inherently multidimensional. Latency affects dependency chains and exposed message startup costs. Bandwidth determines transfer time once communication is on the critical path. Jitter perturbs both average behavior and tails, which is particularly relevant in synchronous training where the slowest participant drives global progress.

Existing methods each capture only part of this landscape. Direct hardware experiments provide indispensable baselines but are limited to available fabrics and costly for broad what-if exploration. Packet-level simulation offers richer controls but becomes infeasible when sweeping multiple network dimensions across production-scale traces. Analytical or reduced-order models can be efficient, yet they often rely on abstractions that are difficult to validate on realistic AI executions~\cite{astra_sim2_won, chakra_sridharan}. What is missing is a validated approach that makes large sensitivity studies feasible while preserving the workload semantics that determine the answer.

This paper addresses this gap by using realistic traces at scales of up to 4,096 GPUs to analyze how latency, bandwidth, and jitter influence end-to-end performance. The main result is a set of latency-bandwidth curves for representative AI workloads, plotted alongside current fabric technologies. These curves identify low-sensitivity regions, transition points, and configurations where runtime increases sharply under the recorded schedule and LogGP model.

To make this study feasible at production scale, we develop a compositional trace analysis methodology grounded in linear programming. Rather than building a single monolithic model for an entire execution, we decompose the workload into its constituent collective operations, analyze each unique communication structure independently, and compose the results at the program level. The method exploits a structural property of AI training: collectives within an iteration are drawn from a small set of distinct signatures, and operations on each CUDA stream execute sequentially. Composition preserves the traced stream order and measured inter-collective compute gaps. This decomposition reduces analysis costs by up to two orders of magnitude. We validate the method against runs on real hardware, monolithic linear programming (LP) solutions, and LogGOPSim~\cite{loggopsim_hoefler} simulations on the studied configurations.

The primary contributions of this work are as follows:
\begin{enumerate}[leftmargin=*]
    \item We provide a large-scale, trace-grounded characterization of AI network sensitivity across realistic workloads, fabrics, and scales up to 4,096 GPUs. The complete dataset contains more than 2\,TB of trace data, and the trace inputs used in this study are publicly available.
    We separate network latency, bandwidth, and jitter, and identify the regions where each dominates as the bottleneck.
    \item We develop and validate a methodology that makes these studies feasible at production size. It combines compositional, linear programming based, trace analysis with signature-based caching and is designed to preserve the regime distinctions and sensitivity classifications that determine provisioning decisions.
    \item We derive latency-bandwidth performance envelopes that expose low-sensitivity regions and performance cliffs. These envelopes provide model-based provisioning guidance and enable direct comparison with modern fabrics.
\end{enumerate}

%% file: Sources/motivation.tex
\section{Background}\label{sec:background}

This section introduces the concepts and tools that underpin the methodology presented in Section~\ref{sec:design}.

\subsection{NCCL and Collective Communication in AI Training}\label{sec:bg-nccl}

Modern distributed AI workloads rely on collective communication to synchronize gradients, broadcast parameters, and redistribute activations across GPUs. On NVIDIA hardware, these collectives are typically implemented by the NVIDIA Collective Communications Library (NCCL)~\cite{nccl_nvidia, hu2025demystifying}, which provides optimized routines for collectives and point-to-point operations.

NCCL divides the execution of collectives into multiple \emph{channels}, which execute the collective's algorithm (typically Ring or Tree) in parallel. Within a single GPU kernel launch, channels map to distinct thread blocks, and, depending on the hardware topology, they can either share a single physical NIC or be routed across multiple NICs to maximize inter-node bandwidth. Within a channel, communication proceeds in a pipelined fashion: data is divided into \emph{slices}, and consecutive slices overlap their send, receive, and optional reduce operations across pipeline stages. The slice size, number of pipeline steps, and protocol (Simple, LL, or LL128) are determined at runtime based on the message size and hardware topology~\cite{hu2025demystifying}.

A training iteration typically issues hundreds of collective calls. For instance, in data-parallel training with the distributed optimizer~\cite{zero_rajbhandari}, each backward-pass layer produces a ReduceScatter of gradients followed by an AllGather of updated parameters, yielding $O(n_{\text{layers}})$ collective invocations per iteration. These collectives are launched on CUDA streams, and operations within a single stream execute sequentially. Different streams may execute concurrently, enabling overlap between communication and computation.


\subsection{Execution Traces and the GOAL Format}\label{sec:bg-traces}

Analyzing the network sensitivity of a full training iteration requires a representation that captures not only the communication operations but also their dependencies, timing, and concurrency structure. We build on the tracing and GOAL~\cite{goal_hoefler} generation pipeline introduced in ATLAHS~\cite{atlahs_bonato}.

Starting from NVIDIA Nsight Systems~\cite{nsight_systems} profiling traces of a real training run, the generator reconstructs each NCCL collective by parsing NVTX markers, kernel metadata, and NCCL runtime logs. For each collective instance, the generator identifies the algorithm (Ring~\cite{RING} or Tree~\cite{BARNETT1995191}), protocol (Simple, LL, LL128), number of channels, data size, slice configuration, and the actual ring or tree topology for each channel. It then emits a GOAL~\cite{goal_hoefler} schedule, represented as a directed acyclic graph (DAG) of \emph{send}, \emph{recv}, and \emph{calc} operations connected by dependency edges. Operations are assigned to the recorded streams and NIC paths.

The GOAL format was originally designed for MPI programs~\cite{loggopsim_hoefler, characterizing_influence_hoefler}. ATLAHS applied it to NCCL collectives by mapping operations on each CUDA stream to a distinct GOAL compute stream. This preserves the parallelism and inter-channel NIC sharing that characterize multi-channel collectives. Sends from multiple streams are serialized when they share the same modeled scale-out injection path. Computation costs for reduction and copy primitives are derived from per-operation hardware measurements collected with NPKit~\cite{npkit} at the target cluster and from event timestamps in the traces.

\subsection{The LogGP Model}\label{sec:bg-loggp}

The LogGP model~\cite{logp_culler, loggp_alexandrov} characterizes point-to-point communication using a small set of parameters: the \emph{latency}~$L$ (time for the first byte to traverse the network), the \emph{gap}~$g$ (minimum interval between consecutive message injections at the NIC), the \emph{gap per byte}~$G$ (the inverse of per-link bandwidth, determining transfer time for long messages), the \emph{overhead}~$o$ (fixed CPU cost per send or receive operation), and the number of processors~$P$. The time to deliver a message of $m$ bytes from send initiation to receive completion is:
\begin{equation}\label{eq:loggp-msg}
    T_{\text{msg}} = L + (m-1) \cdot G + 2o
\end{equation}

LogGOPSim (LGS)~\cite{loggopsim_hoefler} is a message-level simulator that replays GOAL schedules under the LogGP model, and reports the runtime of the full schedule. LGS is one of the fastest available trace simulators (minutes for traces with millions of operations) and has been validated for MPI workloads on HPC clusters~\cite{llamp_shen, noise_in_the_clouds_de_sensi, characterizing_influence_hoefler}. We use LGS as a validation baseline for the LP-based predictions.

\subsection{LP-Based Sensitivity Analysis}\label{sec:bg-lp}

LLAMP~\cite{llamp_shen}, developed for MPI HPC applications, introduced the idea of exposing LogGP model parameters as LP variables rather than fixed constants. Given a GOAL dependency graph $G = (V, E)$, the LP assigns a start-time variable $s_v$ to every vertex $v \in V$ and encodes the LogGP timing constraints as inequalities. For a send-receive pair with message size $m$ bytes between ranks on different nodes:
\begin{equation}\label{eq:recv-constr}
    s_{\text{recv}} \;\geq\; s_{\text{send}} + L + m \cdot G + o
\end{equation}
When $L$ or $G$ is a decision variable, minimizing the makespan $T = \max_v s_v$ yields a one-dimensional optimal value function, such as $T^*(L)$, that is piecewise affine:
\begin{equation}\label{eq:piecewise}
    T^*(L) = \max_{k=1,\ldots,K} \bigl( C_k + \lambda_k \cdot L \bigr)
\end{equation}
Each piece $(C_k, \lambda_k)$ corresponds to a distinct critical path. The slope $\lambda_k$ counts the number of inter-node messages on that path and represents the \emph{latency sensitivity}: the additional runtime per unit of added latency. Analogous formulations hold when the bandwidth parameter $G$ is the decision variable, yielding bandwidth sensitivity.

The piecewise-affine structure follows from LP duality: the optimal value of a parametric LP is a convex piecewise-affine function of its right-hand-side parameters. For a one-dimensional sweep, the active slopes can be obtained from the LP dual variables. To visualize the combined parameter space, we evaluate the independently derived latency and bandwidth envelopes on a common finite grid and classify the dominant effect relative to the baseline. These maps summarize the separate one-dimensional sweeps; they are not solutions of a joint parametric LP.

\paragraph{Limitations of the existing approach}
LLAMP was effective in performing sensitivity analysis. However, it was designed for MPI workloads on traditional HPC workloads, where each rank typically corresponds to a single CPU process on a single node. Applying LP-based analysis to NCCL workloads on multi-GPU clusters raises three challenges that the LLAMP toolchain does not address:

\begin{enumerate}[leftmargin=*]
    \item \textbf{Scale.}
    A full training iteration produces dependency graphs with billions of vertices which would imply billions of constraints. The resulting monolithic LP exceeds the practical limits of current solvers for the kind of parametric sweeps needed for sensitivity studies (hundreds of parameter configurations across multiple network dimensions).

    \item \textbf{NIC injection serialization.}
    NCCL multi-channel collectives simultaneously inject messages from multiple channels using a shared NIC. This results in contention, which is not present in single-process-per-node MPI traces, where each rank typically has dedicated NIC access. Without explicit modeling of NIC serialization, the LP underestimates the bandwidth-bound component of runtime.

    \item \textbf{Scale-up vs. scale-out communication.}
    A node typically hosts multiple GPUs (e.g., four or eight) connected by high-bandwidth scale-up interconnects such as NVLink or InfinityFabric, with latencies of a few hundred nanoseconds and bandwidths exceeding 600\,GB/s. Inter-node transfers traverse the scale-out network (e.g., InfiniBand~\cite{ib2022}, Slingshot~\cite{Desensi2020}, or RoCE) with latencies instead in the microsecond range and per-link bandwidths of 25--200\,GB/s \cite{gpugpuinterconnect}. The network parameters under study ($L$, $G$) only model the scale-out network. An LP formulation that treats all communication uniformly will conflate the two regimes, producing sensitivity estimates that do not reflect the actual network dependence. In this work, we extend the approach proposed by LLAMP and differentiate between scale-up and scale-out networks (both during tracing and in the LP formulation).
\end{enumerate}

Section~\ref{sec:design} describes how we address all three challenges.

%% file: Sources/impl.tex
\section{Methodology}\label{sec:design}

We present a compositional approach, called \emph{Composite LP}, to network sensitivity analysis
of distributed AI workloads.
Given an execution trace of NCCL collective operations,
our method produces one-dimensional piecewise-affine functions
$T(L)$ and $T(G)$ that characterize how the program's
end-to-end runtime responds to changes in network latency~$L$
or bandwidth parameter~$G$. We combine these one-dimensional
envelopes on a finite grid to visualize the latency-bandwidth
parameter space.

\subsection{From Traces to Per-Collective LPs}\label{sec:per-coll-lp}

For each unique collective signature $\sigma$ (described in Sec.~\ref{sec:decomp}), we construct and solve an LP. The construction proceeds in three steps.

\paragraph{Step 1: GOAL generation}
From the signature parameters, we generate a standalone GOAL file
that encodes the collective's communication pattern.
For instance, for a Ring AllReduce on $n$ ranks with $c$ channels, this includes
the reduce-scatter phase ($n{-}1$ steps of receive-reduce-send per channel) followed by the allgather phase ($n{-}1$ steps of receive-copy-send per channel), with explicit per-slice pipelining and inter-channel parallelism on distinct compute streams.

\paragraph{Fixed NCCL schedule}
The analysis replays the NCCL algorithm, protocol, channel count, chunking, and channel topology recorded in the trace. These choices remain fixed while $L$ or $G$ is varied. The method therefore isolates network-parameter sensitivity for an observed schedule; it does not model NCCL selecting a different schedule for another fabric.

\paragraph{Step 2: Dependency graph and LP}
The GOAL file is then parsed into a dependency graph, and the graph is converted to an LP following the formulation of Equation~\eqref{eq:recv-constr}. We extend the LP formulation from Section~\ref{sec:bg-lp} with two refinements required for multi-GPU clusters:

\begin{itemize}
    \item \textbf{Scale-up/Scale-out distinction.}
    For rank pairs running on GPUs on the same node (connected via a scale-up network),
    the latency and bandwidth parameters are fixed constants
    ($L_{\text{scale-up}}$, $G_{\text{scale-up}}$) rather than LP variables.
    This ensures that the sensitivity analysis reflects changes to the
    scale-out fabric only, without conflating scale-up transfers
    that are unaffected by network reconfiguration.

    \item \textbf{NIC injection serialization.}
    In multi-channel collectives, sends from the same source rank to the same
    destination may share a physical NIC injection path.
    We model this contention by introducing ordering constraints between consecutive sends in each per-destination chain:
    \begin{equation}\label{eq:nic}
        s_{\text{send}_{j+1}}^{(r,d)} \;\geq\; s_{\text{send}_j}^{(r,d)}
        + m_j \cdot G + o
    \end{equation}
    where $s_{\text{send}_j}^{(r,d)}$ is the $j$-th send from rank $r$
    to destination $d$, and $m_j$ is the message size.
    Without these constraints, the LP would allow simultaneous injection
    of all channels' messages and underestimate the bandwidth-bound
    runtime component.
\end{itemize}

\paragraph{Step 3: Parametric sweep and piece extraction}
We sweep the decision variables (e.g., $L$ from 0 to $L_{\max}$)
in discrete steps.
At each point, the LP is re-optimized using dual simplex warm-starting:
after tightening the lower bound on~$L$, the previous basis remains dual
feasible, and re-optimization requires a much smaller number of pivots.
From each solve, we extract the reduced cost of the selected decision variable,
which gives the slope of the currently active piece.
The full set of pieces $\{(\lambda_k, C_k)\}$ is collected across
the sweep in a single pass, without requiring separate sensitivity
analysis.

When bandwidth sensitivity is analyzed, we leave $G$ as an LP decision variable
instead of $L$ and sweep $G$ from a high-bandwidth to a low-bandwidth
regime.
For the latency-bandwidth maps, we evaluate the separate $T(L)$ and
$T(G)$ envelopes on a common two-dimensional grid. At each point, we
compare the changes relative to the baseline and classify whether
latency, bandwidth, or both have a substantial effect. The maps are
derived summaries of the one-dimensional sweeps, rather than a joint
two-parameter LP or an exact enumeration of parametric regions.

\subsection{Decomposition via Collective Signatures}\label{sec:decomp}

The per-collective LP described above is exact for a single collective
but would need to be repeated for each of the hundreds of collectives
in a training iteration. We observe that these collectives are drawn
from a small set of \emph{collective signatures}.
A signature $\sigma$ is defined by the tuple:
\begin{equation}\label{eq:signature}
    \sigma = \bigl(\textit{type},\; \textit{algo},\; \textit{proto},\;
    d,\; n,\; c,\; b,\; \tau \bigr)
\end{equation}
where \textit{type} $\in \{\text{AllReduce}, \text{AllGather},
\text{ReduceScatter}, \ldots\}$ is the collective type,
\textit{algo} $\in \{\text{Ring}, \text{Tree}\}$ is the algorithm,
\textit{proto} $\in \{\text{Simple}, \text{LL}, \text{LL128}\}$ is the
NCCL protocol, $d$ is the data size in bytes, $n$ is the number of ranks,
$c$ is the number of channels, $b$ is the slice size in bytes, and
$\tau$ is the ring or tree topology (the permutation defining each rank's
predecessor and successor in each channel).

Two collective instances with the same signature produce identical
dependency graphs and therefore identical piecewise-affine functions
$T_\sigma(L)$.
This property follows from the deterministic mapping between signature
parameters and GOAL structure: the number of pipeline stages, message
sizes, and inter-rank dependencies are fully determined by~$\sigma$.

The number of unique signatures in a training iteration is bounded
by the model architecture, not by the number of collective invocations.
In data-parallel training, each layer produces collectives whose data
size is determined by the parameter tensor shape. The algorithm,
protocol, and topology are selected identically for tensors of the
same size.
In a Llama~8B training iteration with 202 collectives per GPU, we find
only 15 unique signatures (12 Ring-Simple collectives of varying
data size plus 3 negligibly small Tree-LL scalar reductions).

\begin{figure*}[!t]
\centering
\includegraphics[width=1\linewidth]{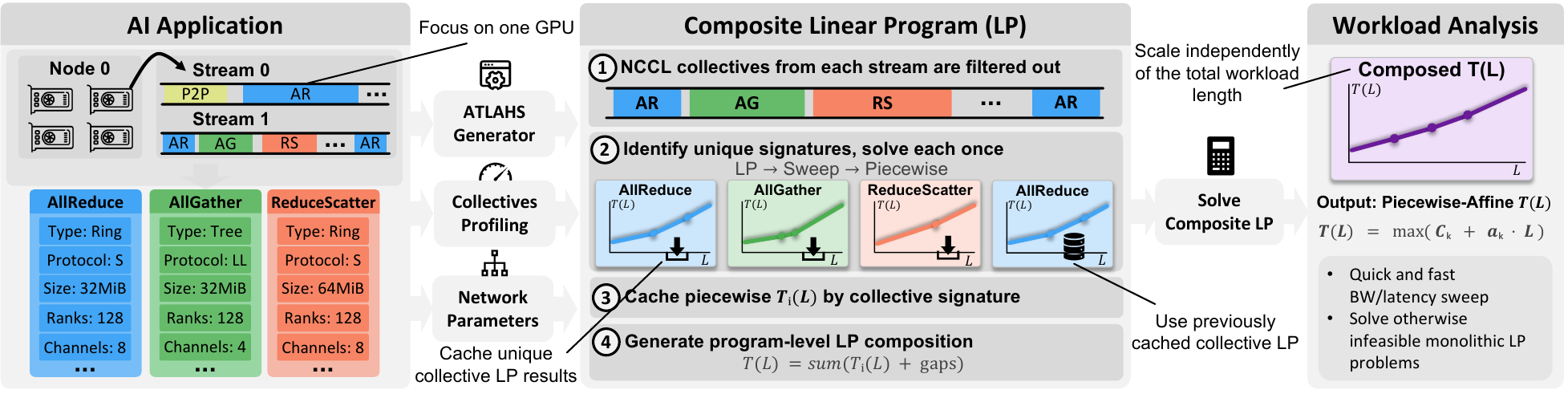}
\caption{Compositional sensitivity analysis pipeline.}
\label{fig:pipeline-overview}
\end{figure*}

\subsection{Program-Level Composition}\label{sec:compose}

Given the per-collective results $T_{\sigma_i}(L)$ for each operation
$i = 1, \ldots, N$ in the program's execution order, plus the
inter-collective compute gaps $g_i$ measured from the trace, we
formulate a lightweight \emph{program-level LP} that composes
individual collective sensitivities into a program-level
$T_{\text{prog}}(L)$.

The decomposition uses two structural properties
of NCCL workloads. First, collectives on each CUDA stream
execute sequentially: the NCCL runtime issues collectives in program
order, and each collective completes before the next one on the same
stream begins. This sequential ordering means that the finish time
of one collective and the start time of the next are linked by a
precedence constraint. Second, the compute gaps between collectives
are measured from the trace and remain fixed during a network-parameter
sweep. The trace conversion obtains cross-stream constraints in a
best-effort manner by merging non-overlapping regions, following
ATLAHS~\cite{atlahs_bonato}. Within each collective, the GOAL graph
retains NCCL's channel-level send, receive, and compute operations,
per-slice pipeline dependencies, scale-up and scale-out distinction,
NIC assignment, and injection serialization.

The program LP introduces, for each stream, variables $s_i$ (start time) and $f_i$
(finish time) for each operation, a free variable $L$, and a makespan
variable $T$:
\begin{alignat}{2}
    f_i &\;\geq\; s_i + C_k + \lambda_k \cdot L
    &\quad& \forall\, k \in \text{pieces}(\sigma_i),\;
            \forall\, i \in \mathcal{C}
    \label{eq:compose-coll} \\
    f_i &\;\geq\; s_i + g_i
    &\quad& \forall\, i \in \mathcal{G}
    \label{eq:compose-gap} \\
    s_{i+1} &\;\geq\; f_i
    &\quad& \forall\, i = 1,\ldots,N{-}1
    \label{eq:compose-seq} \\
    T &\;\geq\; f_N
    &\quad& \label{eq:compose-obj}
\end{alignat}
where $\mathcal{C}$ is the set of collective operations,
$\mathcal{G}$ is the set of compute gaps, and
$\text{pieces}(\sigma_i)$ denotes the piecewise-affine pieces
extracted for collective $i$'s signature.

The number of variables in this LP is $2N + 2$ (two per operation
plus $L$ and $T$), and the number of constraints is
$N + \sum_{i \in \mathcal{C}} |\text{pieces}(\sigma_i)|$,
which in practice amounts to a few hundred for a full training
iteration.

\subsection{Signature-Based Caching}\label{sec:cache}

Because the per-collective LP depends only on the signature $\sigma$,
results can be cached and reused for later identical collectives and across runs.
The cache key is a deterministic hash of all signature fields, so reuse
is valid only when the collective type, algorithm, protocol, message size,
rank count, channel count, slice size, and ring or tree topology match.

Caching provides two levels of amortization:
\begin{enumerate}[leftmargin=*]
    \item \textbf{Within a trace:}
    The LP solve cost is proportional to the number of unique
    signatures $U$, not the total number of collective invocations $N$.
    For Llama~8B, $U = 15$ while $N = 202$, so the per-collective
    solve phase touches each signature once and reuses cached results
    for the remaining 187 invocations.
    \item \textbf{Across traces:}
    Different training configurations can share identical signatures
    for corresponding collectives.
    A cache populated from one trace can be reused when analyzing
    another trace only when all signature fields and fixed model
    parameters match.
\end{enumerate}

Moreover, the $U$ per-signature solves can be executed in parallel since each one operates on an independent LP instance with no shared state.

\subsection{Complexity and Scalability}\label{sec:complexity}

\begin{table}[h]
\centering
\caption{Scaling comparison for a trace with $N$ collective instances,
$U$ unique signatures, dependency graph $G=(V,E)$ for the monolithic
trace, and per-signature subgraphs $G_\sigma=(V_\sigma,E_\sigma)$.
Composite entries give the largest independent signature; its program LP has $O(N)$ variables.}
\label{tab:complexity}
\scriptsize
\setlength{\tabcolsep}{3pt}
\begin{tabular}{@{}lcc@{}}
\toprule
& \textbf{Monolithic LP} & \textbf{Composite LP} \\
\midrule
LP variables & $O(|V|)$ & $O(\max_\sigma |V_\sigma|)$ \\
LP constraints & $O(|V| + |E|)$ & $O(\max_\sigma (|V_\sigma| + |E_\sigma|))$ \\
Memory & $O(|V| + |E|)$ & $O(\max_\sigma (|V_\sigma| + |E_\sigma|))$ \\
\bottomrule
\end{tabular}
\end{table}

Table~\ref{tab:complexity} summarizes the scaling of the monolithic
and compositional approaches.
The critical difference is that the monolithic approach builds a
single LP over the full dependency graph $G=(V,E)$, whose size
grows with the total number of collective instances, while the
compositional approach solves $U$ independent LPs, one per unique
signature, each over a subgraph $G_\sigma$, and composes the
results in a program LP with $O(N)$ variables.
Since $U \ll N$ in the reported Llama~8B case (15 vs.\ 202;
the ratio grows further with more training iterations), the
compositional approach is faster and less memory intensive by several
factors as shown in Section~\ref{sec:evaluation}, an improvement that
enables the analysis of much larger traces.

\begin{figure*}[!t]
\centering
\includegraphics[width=1\linewidth]{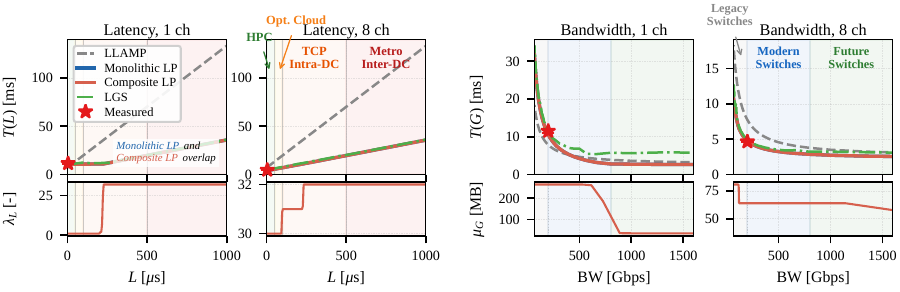}
\caption{Single AllReduce collective of 128MiB running on 16 nodes (64 GPUs) varying the number of NCCL channels.}
\label{fig:val-single}
\end{figure*}

Fig.~\ref{fig:pipeline-overview} illustrates the end-to-end workflow. Starting from a single GPU's perspective, the ATLAHS generator decomposes each CUDA stream into a sequence of collective operations and profiles their signatures (step~\textcircled{1}). Unique signatures are solved independently via LP sweeps (step~\textcircled{2}), and results are cached by signature hash (step~\textcircled{3}). The program-level Composite LP (step~\textcircled{4}) then assembles the cached piecewise-affine functions with the traced compute gaps to produce the final $T(L)$ curve. Because the Composite LP has only $O(N)$ variables, the sweep over hundreds of latency or bandwidth configurations completes in under a second, enabling the multidimensional parameter studies presented in the following sections.

\section{Trace Dataset}\label{sec:traces}

A central goal of this paper is to ground the sensitivity analysis in realistic workloads. To this end, we collect a set of execution traces spanning different model architectures, parallelism strategies, hardware platforms, and scales. The trace inputs and their checksums are available at \url{http://storage2.spcl.ethz.ch/traces/ai/}.

The traces are collected on two supercomputer platforms: the Alps~\cite{cscs_alps} system at CSCS (NVIDIA GH200, Slingshot interconnect~\cite{Desensi2020}), which provides up to 200\,Gbps of scale-out bandwidth, and a large-scale Microsoft Azure GPU cluster, named ND GB200 v6, which provides up to 800\,Gbps. The Azure cluster comprises GB200 nodes, each featuring a dual-socket Grace CPU with 144 cores (72 per socket) and 1.7\,TB of RAM. Each node contains four NVIDIA B200 GPUs with 185\,GB of HBM3e each. GPUs are interconnected via NVLink both within a node and within racks, where each rack houses 18 nodes (72 GPUs). Each GPU is equipped with a dedicated NIC enabling GPUDirect RDMA~\cite{10.1109/ICPP.2013.17} over the backend network, which is organized as a rail-optimized fat tree. The dataset comprises the following workloads:

\paragraph{Llama 3.3}
Meta's Llama~3 family~\cite{llama3_grattafiori} represents the current generation of open-weight large language models. We trace Llama~3.3 training at multiple scale points using Megatron-LM~\cite{megatron_shoeybi} with data parallelism (DP), tensor parallelism (TP), and pipeline parallelism (PP).

\paragraph{Grok}
The Grok workload uses xAI's 314-billion-parameter Mixture-of-Experts (MoE) architecture~\cite{grok1_xai,shazeer2017outrageouslylargeneuralnetworks}, which adds expert-parallel All-to-All communication to the standard DP/TP/PP pattern. We trace Grok at the largest scale in our dataset (4,096 GPUs), producing over 2\,TB of trace data containing more than 16 billion GOAL vertices and 20 billion dependencies.

\paragraph{vLLM inference (DeepSeek, Llama)}
To extend the study beyond training, we include inference traces collected with the vLLM serving engine~\cite{vllm_kwon} running DeepSeek~\cite{deepseek_v3} and Llama models. Inference exhibits different communication patterns: shorter bursts, smaller message sizes. It tends to be more memory-bound.

We ran the Grok traces on Azure ND GB200 v6, while all the other workloads on Alps.


\section{Validation}\label{sec:validation}

Before using the Composite LP method for large-scale case studies, we validate it against three baselines: (1)~the monolithic LP formulation applied to the full GOAL trace, (2)~LogGOPSim (LGS) simulation, and (3)~hardware measurements. We proceed from controlled single-collective experiments to full training iterations. For the validation runs, we use the Alps supercomputer.

Our LP-based approach produces a \emph{lower bound} on runtime. Real networks introduce effects such as congestion, routing variability, transport state, and protocol overheads that are not captured by the LogGP model~\cite{loggopsim_hoefler}. The LP prediction should therefore be interpreted as the best runtime allowed by the recorded dependency structure and modeled parameters, not as a guarantee for a deployed fabric. A gap from hardware or a more detailed simulator may include these omitted effects as well as measurement and modeling error. The LP identifies performance cliffs forced by the recorded dependency graph under the LogGP abstraction. Modeling all omitted effects at every point would make broad parameter sweeps at thousands of GPUs impractical. Our method trades this detail for the ability to evaluate many candidate network configurations while preserving the recorded program dependencies.

\subsection{Single Collective}\label{sec:val-single}

We begin with a single 128\,MiB AllReduce collective on 64 GPUs across 16 nodes (4 GPUs per node), using the Ring algorithm with NCCL's Simple protocol. We run a forced 1-channel configuration and then also use automatic channel selection, which converges to 8 channels. We do this to demonstrate the feasibility of our toolchain to adapt to different configurations. Fig.~\ref{fig:val-single} shows the latency and bandwidth sensitivity curve $T(L)$ predicted by four methods: LLAMP (the original LP formulation created for HPC apps), Monolithic LP (our extended formulation), our Composite LP approach, and LGS simulation. The hardware measurement, averaged over 30 runs at the cluster's
native network parameters, is shown as a reference point (star
marker). In each panel, the upper plot reports the predicted runtime
as a function of either latency or bandwidth, while the
lower plot reports the corresponding sensitivity coefficient.
$\lambda_L=\partial T/\partial L$ is dimensionless and counts the
inter-node messages on the active critical path. The bandwidth sensitivity
$\mu_G=\partial T/\partial G$ has units of bytes because $G$ is the gap per
byte. For readability, the bandwidth panels place $\mu_G$ against the
equivalent bandwidth $B=1/G$ in consistent units. A derivative taken
directly with respect to $B$ would instead have units of time squared per
byte.

\begin{figure}[t]
\centering
\includegraphics[width=\columnwidth]{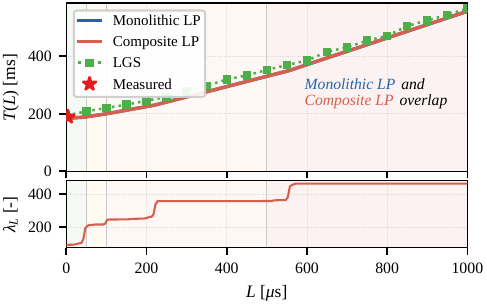}
\caption{Predicted total runtime (top) and latency sensitivity (bottom) of a workload of 20 randomly selected NCCL collectives on 16 nodes (64 GPUs); the Composite LP matches the Monolithic LP result.}
\label{fig:val-mixed}
\end{figure}

To aid interpretation, we annotate the plots with shaded regions corresponding to representative network operating regimes. For latency, we distinguish four regimes: specialized HPC and modern AI fabrics such as Slingshot and InfiniBand HDR/NDR, which typically
operate below $50\,\mu$s~\cite{Desensi2020, ethernet}; commodity cloud interconnects using RDMA, which range from roughly $50$ to $100\,\mu$s~\cite{rdma}; TCP-based protocols,
which introduce software stack overhead and fall in the $100$ to
$500\,\mu$s range~\cite{arrakis}; and metro or inter-datacenter links,
which exceed $500\,\mu$s due to propagation delay \cite{uno, b4_wan}. For bandwidth, we
mark three regimes: current-generation switch bandwidth ($200$ to
$800$\,Gbps), legacy bandwidth (below $200$\,Gbps), and
next-generation or future bandwidth (above $800$\,Gbps) \cite{ethernet}. These boundaries are approximate and serve as visual anchors rather than strict thresholds; we use a consistent color scheme across all figures for ease of cross-referencing.

Three observations stand out. First, the Monolithic LP and Composite LP curves are indistinguishable, confirming that the piecewise extraction faithfully represents the full LP solution in this experiment. Second, LGS and the Monolithic LP closely match throughout the latency sweep, providing a check against an independent simulator. Third, LLAMP underestimates runtime (15\% error at $T(0)$) and overestimates latency sensitivity ($\lambda = 15$ vs.\ $\lambda = 4$), because it treats all communication uniformly without distinguishing scale-up and scale-out networks or modeling NIC serialization and NCCL streams. This result shows why the extensions described in Section~\ref{sec:per-coll-lp} are needed for the studied multi-GPU NCCL collective.

\subsection{Mixed Collectives}\label{sec:val-mixed}

\begin{figure}[t]
\centering
\includegraphics[width=\columnwidth]{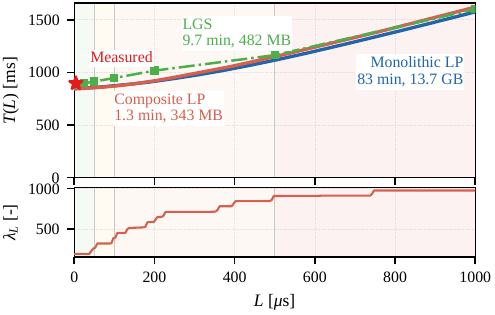}
\caption{Llama~8B training iteration (DP16, 4 nodes, 16 GPUs). Top: runtime $T(L)$; bottom: latency
sensitivity $\lambda_L$. The Composite LP matches the Monolithic LP
while reducing solve time from 83 minutes to under 2 minutes.}
\label{fig:val-llama}
\end{figure}

To further validate the composition methodology, we construct a workload of 20 randomly selected NCCL collectives with data sizes drawn uniformly from 16 to 64\,MiB separated by random compute gaps, executed on 64 GPUs across 16 nodes. Fig.~\ref{fig:val-mixed} compares the Monolithic LP, the Composite LP, LGS, and the hardware wall-clock time. We note that the collectives are separated by a random delay that we intentionally keep short (below 10\% of the total runtime) to clearly analyze the communication patterns.

The Composite LP matches the Monolithic within 0.8\% at $T(0)$ and reproduces the same latency sensitivity ($\lambda = 104$). This confirms that the per-collective decomposition and program-level recomposition preserve the critical-path structure of the full trace in this experiment. LGS reports a higher $T(0)$ (5.3\% above the Monolithic LP), consistent with its stricter NIC serialization model: LGS enforces a single NIC queue per rank across all destinations, while the LP serializes sends only within each per-destination channel group. At the measured hardware point, the LP result is closer to the observed runtime.

\subsection{End-to-End Training: Llama 8B}\label{sec:val-llama}

We now scale the validation to a full Llama~8B training iteration with 16-way data parallelism (4 nodes, 16 GPUs in total), comprising 202 collective operations per GPU. Fig.~\ref{fig:val-llama} compares the Composite LP approach against the Monolithic LP.

The results demonstrate both accuracy and scalability. At each value where the Monolithic LP is feasible, the Composite prediction matches within 0.7\% at $T(0)$. The full Monolithic LP (2.3 million variables, 83 minutes solve time, 14\,GB of memory) confirms the Composite prediction of $T(0) = 841.8$\,ms vs.\ the Monolithic's 847.6\,ms.

The scalability advantage is already significant at this small scale. Composite LP solves all 202 collectives in 45\,seconds (77\,s including cold-start LP solves for all 12 unique Ring-Simple signatures) using at most 343\,MB of memory. The monolithic approach requires 83 minutes and 14\,GB. If the 12 per-signature LP solves are run in parallel, with 8 cores, the solve phase completes even faster in just 22\,seconds.

\subsection{Validation Across Scale}\label{sec:val-scaling}

We also evaluate Composite LP and LGS across nine Grok scale points, from 16 to 4,096 GPUs. At 4\,$\mu$s, in line with the inter-node latency on Alps, Fig.~\ref{fig:grok-scaling} lets us validate the model at different GPU counts against both LGS and hardware. Across the eight scale points whose exports contain all ranks, Composite LP and LGS have mean errors of about 9\% and 3\%, respectively. At 4,096 GPUs, both remain within 10\% of the available hardware reference.

The remaining panels compare Composite LP and LGS for higher hypothetical latencies. Raising inter-node latency from 4\,$\mu$s to 1\,ms changes the predicted runtime by less than 0.3\% at 16 GPUs. At 4,096 GPUs, the Composite LP prediction rises from about 11 to 31\,s, while LGS rises from about 12 to 33\,s. The two methods therefore show the same increase in latency sensitivity as the job grows.



%% file: Sources/evaluation.tex
\section{Evaluation}\label{sec:evaluation}

Having validated the compositional methodology on the configurations in Section~\ref{sec:validation}, we now apply it to the trace dataset introduced in Section~\ref{sec:traces}. At production scale, the Monolithic LP is infeasible, and repeated LGS runs across broad parameter sweeps are costly.

\begin{figure}[H]
\centering
\includegraphics[width=\columnwidth]{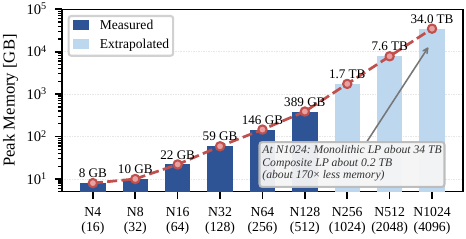}
\caption{Monolithic LP's peak memory footprint for Grok as a function of GPU count. Dark blue bars are measured; light blue bars are extrapolated. At 4,096 GPUs the monolithic formulation would require over 34\,TB.}
\label{fig:memory}
\end{figure}

Fig.~\ref{fig:memory} quantifies this limitation for Grok. We measure the peak memory footprint of the Monolithic LP at scale points from 4 to 128 nodes (16 to 512 GPUs), and extrapolate the larger configurations starting at 1,024 GPUs through 4,096 GPUs. Memory footprint grows superlinearly with GPU count: 8\,GB at 16 GPUs, 59\,GB at 128, 146\,GB at 256, and 389\,GB at 512 GPUs. Extrapolating to the full 4,096-GPU configuration yields an estimated 34\,TB. The Composite LP, by contrast, solves each collective signature independently; its memory footprint is bounded by the largest individual collective and peaks at around 200\,GB for 4,096 GPUs. Composite LP takes around 4 hours for this case, while we estimate that Monolithic LP would take several weeks. For this reason, all results in this section use Composite LP, with LGS simulation and hardware measurements as cross-checks where feasible.

\begin{figure*}[!t]
\centering
\includegraphics[width=1\linewidth]{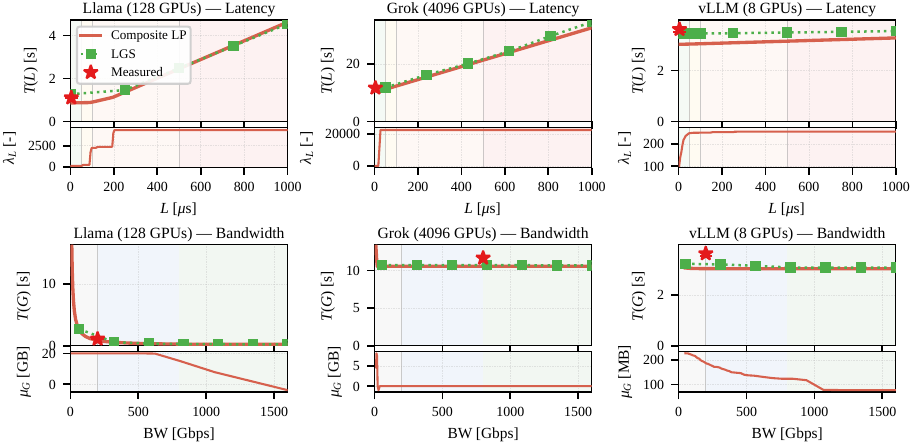}
\caption{Model-derived latency (top) and bandwidth (bottom) sensitivity for Llama~70B (128 GPUs), Grok (4,096 GPUs), and vLLM Llama-3.1-8B (8 GPUs). Each panel shows the predicted runtime and corresponding sensitivity coefficient ($\lambda_L$ for latency, $\mu_G$ for bandwidth). Available hardware measurements and LGS predictions are overlaid. The bandwidth hardware points use each system's native scale-out rate: 200\,Gbps on Alps for Llama and vLLM, and 800\,Gbps on Azure ND GB200 v6 for Grok.}
\label{fig:grok-llama}
\end{figure*}

\begin{figure*}[!t]
\centering
\includegraphics[width=1\linewidth]{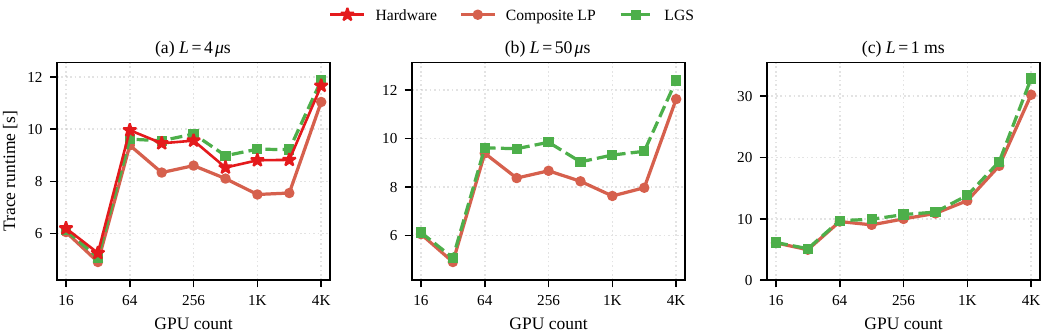}
\caption{Grok trace runtime from 16 to 4,096 GPUs, with four GPUs per node, at inter-node latencies of (a)~4\,$\mu$s, (b)~50\,$\mu$s, and (c)~1\,ms. The measured runtime at the native hardware settings is shown in (a).}
\label{fig:grok-scaling}
\end{figure*}

\subsection{Latency and Bandwidth Sensitivity}\label{sec:eval-lb}

Fig.~\ref{fig:grok-llama} presents the latency and bandwidth
sensitivity profiles for three representative workloads, spanning AI training and inference: Llama~70B training on 128 GPUs, Grok training on 4,096 GPUs, and vLLM Llama~8B inference on 8 GPUs. The top row sweeps inter-node latency $L$ from 0 to 1\,ms; the bottom row sweeps bandwidth from 10 to 1,600\,Gbps.

\paragraph{Latency}
The two training workloads respond to latency inflation in qualitatively different ways. Llama~70B exhibits a piecewise structure with two key jump points near $L = 100$\,$\mu$s and $L = 200$\,$\mu$s: below the smaller threshold, the runtime is nearly flat, while above it, latency begins to dominate. The $\lambda_L$ subplot confirms this, rising from near zero to approximately 2,500 and then 5,000 as latency grows. This transition reflects the point at which inter-node communication displaces computation or bandwidth from the critical path.

Grok at 4,096 GPUs shows only a small low-sensitivity region below $50\,\mu$s. From that point, its $\lambda_L$ is approximately 22,500 across the rest of the sweep, meaning that latency is on the critical path for most evaluated points. At 1,024 nodes, ring traversals cross enough inter-node hops that communication becomes exposed even at relatively low latency values.

For vLLM, latency sensitivity similarly rises to $\lambda_L \sim 60$ for $L \lesssim 50\,\mu$s, beyond which it stays relatively flat. However, the overall effect of latency variation on end-to-end runtime is limited compared with the training workloads investigated here.

\paragraph{Bandwidth}
Bandwidth sensitivity tells a different story. The Llama~70B bandwidth curve reveals a sharp transition below 200\,Gbps, where reducing bandwidth causes runtime to increase rapidly. Above 800\,Gbps, additional bandwidth yields diminishing returns, as computation and latency-bound paths dominate the critical path. The bandwidth sensitivity, $\mu_G$, confirms this structure. This study keeps compute performance fixed while varying network latency and bandwidth. Faster compute subsystems could expose more communication and therefore change the sensitivity envelope; these sweeps should not be read as predictions for future accelerators.

Grok, on the other hand, exhibits near-zero bandwidth sensitivity, as its runtime is largely driven by compute costs. For inference and vLLM~8B, we observe a similar trend: bandwidth matters little, and the critical path carries only a very small byte volume.

\begin{figure}[t]
\centering
\includegraphics[width=\columnwidth]{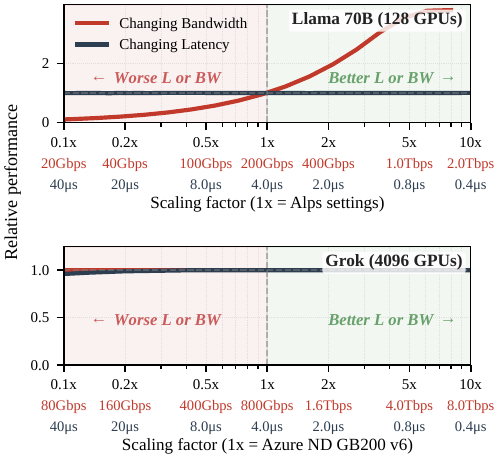}
\caption{Latency and bandwidth sensitivity for Llama 70B and Grok focusing on AI/HPC specialized cluster network parameters.}
\label{fig:dual}
\end{figure}

\paragraph{Current Data Center Networks}
However, the latency range swept in Fig.~\ref{fig:grok-llama}
extends well beyond what modern datacenter fabrics deliver.
Production Slingshot and InfiniBand networks typically operate in the
1--50\,$\mu$s latency range at 200--800\,Gbps bandwidth. When the analysis is restricted to this window, the picture changes considerably, as
shown in Fig.~\ref{fig:dual}. For both Llama and Grok, latency
variations within this range have negligible effect on runtime.
Bandwidth sensitivity, however, diverges: Grok remains largely
insensitive, whereas Llama exhibits high sensitivity, particularly
to bandwidth reductions. At 20\,Gbps, Llama's iteration time
increases by nearly 30$\times$ relative to the baseline. 

In Figs.~\ref{fig:dual} and~\ref{fig:cost_modelling}, relative performance is $P_{\mathrm{rel}}=T_{\mathrm{base}}/T$, where $T_{\mathrm{base}}$ is the modeled runtime at the corresponding native bandwidth or latency baseline. Thus, $P_{\mathrm{rel}}=1$ matches the baseline and $P_{\mathrm{rel}}=0.01$ denotes a $100\times$ slowdown.

Within the latency and bandwidth ranges characteristic of current data center fabrics, bandwidth is the dominant network parameter for the studied Llama configuration. In the model, latency variation within the low-microsecond range has little effect for these workloads. The wider sweep also shows that latency becomes more important for configurations that cross metropolitan or inter-data-center distances. These observations are workload-specific and should be evaluated together with the assumptions in Section~\ref{sec:limitations}.

\subsection{Cost-Performance Tradeoff}\label{sec:eval-cost}

\begin{figure}[h]
\centering
\includegraphics[width=\columnwidth]{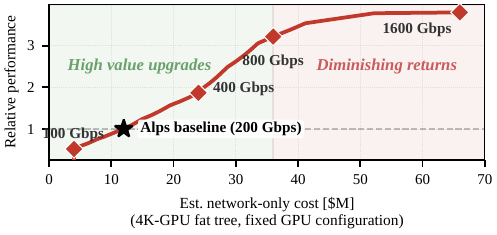}
\caption{Estimated network-only cost versus relative performance for Llama~70B on 128 GPUs. Each point corresponds to a different bandwidth (100--1,600\,Gbps) under fixed topology, radix, NIC count, and GPU configuration. The baseline marks 200\,Gbps, as used by Alps in our experiments.}
\label{fig:cost_modelling}
\end{figure}

The bandwidth sensitivity results do not show how the performance gains compare with the additional network cost. To provide this comparison, Fig.~\ref{fig:cost_modelling} maps bandwidth operating points onto estimated switch and network-adapter costs for a fixed 4,096-GPU fat-tree deployment. The topology, radix, NIC count, and GPU configuration remain fixed while the per-link bandwidth changes.

The plot shows a hyperbolic shape with two distinct regions. On the left, upgrading from 100 to 800\,Gbps delivers large iteration-time reductions per dollar spent. The transition from 100 to 200\,Gbps alone roughly halves the iteration time. On the right, upgrading from 800 to 1,600\,Gbps nearly doubles the network cost while reducing iteration time by a small fraction. 

For the same modeled GPU fleet, the performance gained per unit of network cost depends on the operating point relative to the workload's bandwidth breakpoint. Below that point, bandwidth upgrades produce larger modeled gains. Above it, the gains diminish.

Detailed network cost modeling is beyond the scope of this paper. Our
estimates use prior work and public list prices for switches and network
adapters~\cite{hammingmesh_hoefler, 10.1109/SC.2014.34, fs_nvidia_switch_2026, dell_connectx7_2026}.
They provide a relative network-only comparison under fixed design
assumptions, not a procurement quote or an estimate of total system cost.





\subsection{Jitter}\label{sec:eval-jitter}

The results so far treat network parameters as deterministic: every inter-node message sees the same latency $L$. In production networks, however, some messages may experience elevated delays due to congestion, routing variability, or transient failures~\cite{gpugpuinterconnect,noise_in_the_clouds_de_sensi,netnoisearies}. We use a controlled perturbation to study how delayed messages affect the modeled runtime. Table~\ref{tab:jitter} reports minimum and maximum latency and bandwidth values from previous network characterization studies.

\begin{table}[htbp]
\centering
\caption{Reference network latency and bandwidth values from previous characterization studies.}
\label{tab:jitter}
\resizebox{\columnwidth}{!}{
\begin{tabular}{llllll}

\textbf{System} & \textbf{Min Lat.} & \makecell[l]{\textbf{Max Lat.}\\\textbf{(Increase)}} & \textbf{Max Bw.} & \makecell[l]{\textbf{Min Bw.}\\\textbf{(Decrease)}} \\ \hline
Alps (Slingshot 11) \cite{gpugpuinterconnect}  & 3.64 $\mu$s & 6.48 $\mu$s ($1.78\times$) & 763 Gb/s & 763 Gb/s ($1\times$) \\
LUMI (Slingshot 11) \cite{gpugpuinterconnect}  & 2.31 $\mu$s & 4.62 $\mu$s ($2\times$) & 766 Gb/s & 766 Gb/s ($1\times$) \\
Leonardo (IB HDR) \cite{gpugpuinterconnect}  & 1.9 $\mu$s & 131.86 $\mu$s ($69.4\times$) & 395 Gb/s & 210.11 Gb/s ($1.88\times$) \\
Daint (Aries) \cite{noise_in_the_clouds_de_sensi}  & 1.192 $\mu$s & 610.30 $\mu$s ($512\times$) & 75.24 Gb/s & 37.62 Gb/s ($2\times$) \\
Oracle (RoCEv2) \cite{noise_in_the_clouds_de_sensi} & 1.66 $\mu$s & 49.80 $\mu$s ($30\times$) & 97.53 Gb/s & 2.24 Gb/s ($43.5\times$) \\
AWS (SRD) \cite{noise_in_the_clouds_de_sensi} & 16.79 $\mu$s & 537.28 $\mu$s ($32\times$) & 78.74 Gb/s & 39.37 Gb/s ($2\times$)  \\ 
\end{tabular}
}
\end{table}

We use a controlled two-state model with a nominal latency of 5\,$\mu$s and an elevated latency of 250\,$\mu$s. For an elevated-latency fraction $p$, we interpolate the Composite LP predictions as $T(p)=(1-p)T(5\,\mu\mathrm{s})+pT(250\,\mu\mathrm{s})$. We sweep $p$ from 0 to 50\%.

\begin{figure}[t]
\centering
\includegraphics[width=\columnwidth]{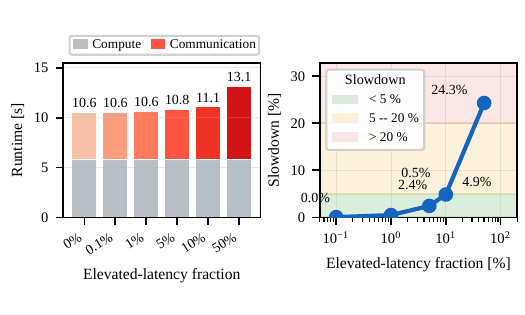}
\caption{Impact of the two-state latency perturbation model on Grok (4,096 GPUs). (a)~Iteration time as a function of the elevated-latency fraction, decomposed into compute and communication. (b)~Slowdown as a function of the elevated-latency fraction.}
\label{fig:jitter}
\end{figure}

Fig.~\ref{fig:jitter} shows the results for Grok. Under this two-state model, small elevated-latency fractions have a limited effect: at $p = 0.1\%$ the slowdown is 0.1\%, and at $p = 1\%$ it remains within 1\%. At $p = 10\%$ the slowdown reaches 5\%, and at $p = 50\%$ it exceeds 24\%.

These results complement the deterministic latency and bandwidth analysis with a controlled trace-based perturbation. They do not represent correlated congestion or a measured tail-latency distribution.

\subsection{Scope and Limitations}\label{sec:limitations}

The reported envelopes are structural lower bounds under the recorded dependency graph and the LogGP abstraction. They do not model topology-specific routing, adaptive routing, background traffic, congestion, transport state, retransmissions, or other fabric effects. These effects can increase runtime and can shift the observed breakpoints. A low-sensitivity region in our model is therefore not a guarantee of operational performance on a loaded system.

The analysis also keeps the traced NCCL schedule and compute behavior fixed. It does not retune the collective algorithm, protocol, channel count, chunking, or topology when network parameters change. Within collectives, the GOAL graph retains channel-level operations, pipeline dependencies, scale-up and scale-out transfers, and modeled NIC serialization. At program level, it preserves CUDA stream order and measured compute gaps, while cross-stream constraints are inferred in a best-effort manner. Changes in accelerator performance, overlap, or schedule selection can therefore produce different envelopes. Future work could dynamically select NCCL algorithms and protocols or add more detailed routing and transport effects. Both extensions would increase analysis time, which is the main tradeoff against the broad sweeps studied here.

ATLAHS currently relies on kernel-level activity and NVTX metadata in Nsight Systems exports. Recent Nsight Systems versions trace CUDA Graphs at graph granularity by default, where the underlying node activities are not collected~\cite{nsight_systems}. This can hide kernels in inference workloads that rely heavily on CUDA Graph replay. Node-level graph tracing is available, but it can add significant profiling overhead. Supporting graph-level inference traces is an important direction for future work.

Finally, the jitter study uses a two-state interpolation and is not a model of bursty or correlated network events. Hardware validation is performed where native measurements are feasible, while the broad and large-scale parameter sweeps remain model-based. The quantitative conclusions apply to the recorded workloads, software choices, and scale points; the toolchain can be rerun for other configurations.

%% file: Sources/related.tex
\section{Related Work}\label{sec:related}

\paragraph{Network performance modeling}
The LogP~\cite{logp_culler} and LogGP~\cite{loggp_alexandrov} families capture point-to-point communication using a small set of abstract cost parameters. Building on these models, LogGOPSim~\cite{loggopsim_hoefler} replays GOAL schedules and has been validated on large-scale MPI workloads~\cite{characterizing_influence_hoefler, noise_in_the_clouds_de_sensi}.
LLAMP~\cite{llamp_shen} reformulates the LogGP constraints as a linear program, enabling analytical sensitivity and tolerance analysis for MPI applications without repeated simulation.
Our work builds on the LP formulation but targets NCCL-based AI workloads and introduces a compositional decomposition that reduces analysis cost by up to two orders of magnitude.

\paragraph{AI training simulation}
Simulators such as ASTRA-sim~\cite{astra_sim2_won} and SimAI~\cite{simai_wang} model distributed training by coupling compute, memory, and network simulators, enabling what-if studies on parallelism strategies and hardware configurations.
Chakra~\cite{chakra_sridharan} proposes an open format to store execution traces of ML workloads and benchmarks and supports trace-driven simulation.
Both frameworks target design-space exploration at the system level; they do not expose one-dimensional parametric network sensitivity as a piecewise-affine function.
Packet-level simulators such as NS-3~\cite{ns3_riley} and htsim~\cite{htsim_raiciu} provide high-fidelity network modeling but are too costly for the multidimensional sweeps required by our study~\cite{towards_million_besta}.
ATLAHS~\cite{atlahs_bonato} bridges these approaches by converting real application traces to GOAL and feeding them to multiple simulator backends; GOAL traces are $2\times$--$11\times$ smaller than corresponding Chakra traces~\cite{atlahs_bonato}. We adopt the ATLAHS tracing infrastructure and extend the analysis layer with LP-based composition.

\paragraph{Network sensitivity and provisioning}
Prior work on network latency sensitivity has focused on MPI workloads.
Underwood et al.~\cite{underwood_latency} and Patke et al.~\cite{patke_latency} inject latency on real hardware to measure application response, an approach that is effective but expensive and limited to available fabrics.
Hoefler et al.~\cite{characterizing_influence_hoefler} study the influence of system noise on HPC applications via simulation.
De~Sensi et al.~\cite{noise_in_the_clouds_de_sensi} characterize network variability in cloud environments.
These efforts examine individual dimensions (latency or noise) on HPC codes.
Our study examines latency, bandwidth, and jitter on large AI workload traces, derives piecewise-affine one-dimensional envelopes, and uses the latency and bandwidth envelopes to construct finite-grid sensitivity maps.

%% file: Sources/conclusions.tex
\section{Conclusion}\label{sec:conclusions}

We presented a trace-grounded study of network sensitivity for AI workloads on supercomputers. By decomposing execution traces into per-collective LP problems and composing them at the program level, we reduced the cost of parametric sensitivity analysis by up to two orders of magnitude compared to monolithic formulations, making multidimensional sweeps over latency, bandwidth, and jitter feasible at large scale, including the 4,096 GPUs test case presented earlier.

The resulting performance envelopes show that, depending on the workload, large regions of the latency-bandwidth space produce near-identical runtime under the model and that transitions to degraded operation can be sharp.
These findings provide provisioning guidance by identifying low-sensitivity regions, performance cliffs, and points of diminishing returns. They should be interpreted together with the lower-bound model assumptions and workload scope discussed in Section~\ref{sec:limitations}.

The execution traces used in this study are available at \url{http://storage2.spcl.ethz.ch/traces/ai/}. They enable the analysis to be repeated for the workloads and scale points studied in this work.

\section*{Acknowledgments}
This work is supported by the European Union’s Horizon Europe under grant 101175702 (NET4EXA), by Sapienza University Grants ADAGIO and D2QNeT (\textit{Bando per la ricerca di Ateneo} 2023 and 2024), and by a research grant from Microsoft Azure. The research was also conducted as part of the FastTrackAI project at the Singapore-ETH Centre, which was established collaboratively between ETH Zurich and the National Research Foundation, Singapore. Additionally, this research is supported by the National Research Foundation, Singapore (NRF), and the Ministry of Digital Development and Information (MDDI) under the AI Visiting Professorship (Award No. AIVP-2025-005). We also thank the Swiss National Supercomputing Center (CSCS) for providing the computational resources used in this work.

ChatGPT 5.4 and Opus 4.6 were used for language editing.